\documentclass{article}
\usepackage{spconf,amsmath,graphicx,hyperref}
\usepackage{tabularx}
\usepackage{booktabs}
\usepackage{comment,color}

\title{Lightweight Systems \\ for Spatial Semantic Segmentation of Sound Scenes}
\name{Mayank Mishra, Paul Magron, Romain Serizel}
\address{Université de Lorraine, CNRS, Inria, LORIA, F-54000 Nancy, France}

\begin{document}
\ninept
\maketitle
\begin{abstract}
Spatial semantic segmentation of sound scenes (S5) jointly performs audio source separation and tagging on a multichannel audio mixture. Therefore, an S5 system can be decomposed into an audio tagging model and a separation model. The baseline S5 system in task~4 of the DCASE 2025 challenge uses M2D and ResUNet models for tagging and separation, respectively. These are large in overall size and therefore not appropriate for use in embedded devices. The main aim of this paper is to assess the performance loss in an S5 system when using lightweight tagging and separation models instead. To this end, we first compare the tagging performance of lightweight audio tagging models, namely PANNs and DyMNs, in terms of label prediction accuracy against the baseline M2D model. Secondly, we assess separation performance in terms of the scale-invariant signal-to-distortion ratio of the baseline ResUNet separator relative to that of a convolutional-based separator using SuDoRMRF. Finally, we combine the tagging and separation models across various configurations and assess the overall performance of the S5 system using the class-aware signal-to-distortion ratio.
\end{abstract}
\section{Introduction}
\label{sec:intro}


Spatial semantic segmentation of sound scenes (S5) is a recent task proposed in the Detection and Classification of Acoustic Scenes and Events (DCASE) Challenge 2025. The task requires creating a system that takes as input a multichannel audio signal containing a mixture of target sources, and returns single-channel, separated sources with their class labels~\cite{yasuda2025}. Applications of S5 include immersive communications, extended reality, and assisted living in smart homes~\cite{yasuda_spatial_2025, virtanen2017computational}.


To develop an S5 system, one approach is to use an audio tagging model and a source separation model sequentially. In general, the two components of the system can be used in any order to perform S5, either using the tagging model first and then performing source separation or vice versa~\cite{Kong2017AJS}. When the tagging model is used first, the source separation is often referred to as \textit{target sound extraction} (TSE)~\cite{Soundsculpt, Wu2024CrossattentionIS, Wang2024SoloAudioTS, Yin2025TextQueriedAS, Kong2026UniversalSS}, as the labels are used as queries to the source separation model. Finally, another approach is also possible, where an end-to-end S5 system can be designed. Such a system then consists of a hybrid model that performs both tagging and separation together~\cite{Bando_2025_t4, Qian_2025_t4}.

A better-performing S5 system is usually obtained by having a better-performing tagger or separator. One way to improve the performance of a deep learning model is to adopt more complex architectures, which usually increases model size. In the case of an S5 system, this effect is compounded as it is made up of two components in different domains, namely a tagger and a separator. This effect is evident in the model sizes of submissions to task 4 of the DCASE 2025 challenge,\footnote{https://dcase.community/challenge2025/task-spatial-semantic-segmentation-of-sound-scenes-results} where the largest S5 system has approximately 1000 times more parameters than the smallest system.

Their large model sizes prevent these systems~\cite{Choi_2025_t4, FulinWu_2025_t4, Morocutti_2025_t4} from being used on certain edge applications, such as wearable devices and devices for home-assisted living, while the smallest ones tend to underperform. There is a need to design S5 systems that are lightweight and can be easily incorporated into such devices while maintaining high performance. In the broader audio research field, several studies have utilized lightweight architectures for audio tagging~\cite{Kong2019PANNsLP, Schmid2023DynamicCN}, speech enhancement~\cite{Fedorov2020TinyLSTMsEN}, and audio and music separation~\cite{ Wang2022LightweightED}. However, none of these mention anything similar to a complete S5 system. A major difficulty in using lighter architectures for an S5 system is that it includes two different components, and making one of them lighter might severely affect the other component, leading to a reduced overall performance, as it was previously observed in hybrid speech enhancement systems~\cite{benslimane2026rttangorealtimedistributedbinaural}.

As an initial contribution towards lightweight S5, we propose leveraging lighter audio tagging and source separation models to build an efficient S5 system. We use two convolution-based audio tagging models~\cite{Kong2019PANNsLP, Schmid2023DynamicCN}, which are computationally less expensive than currently popular transformer-based models. We also propose using a mask-based convolutional architecture as a separator~\cite{tzinis2022compute} and modify it to be able to perform TSE. Finally, we study the impact of changing one of the components with a lighter model in a full S5 setup and also compare its performance with that of the DCASE 2025 baseline system.





\section{Lightweight S5 systems}
\label{sec:methodo}

To develop an S5 system, we follow an approach that is proposed by most S5 systems in the literature, i.e., using the tagger first and then the separator for TSE. Here, we describe the baseline S5 system, the tagging and TSE models used, and their modifications.

\subsection{Baseline system}
\label{sec:methodo-baseline}

The baseline S5 system~\cite{nguyen2025baselinesystemsevaluationmetrics} uses a masked modeling duo (M2D) model as a tagger. M2D is a self-supervised foundation model that learns representations of masked patches from unmasked patches~\cite{10502167}.
M2D can be adjusted for application to any downstream audio classification task, such as environmental sound classification, urban sound classification, or musical genre recognition. In particular, M2D was pretrained using the AudioSet dataset~\cite{AudioSet} labels to perform the audio classification task~\cite{10502167}. In the S5 baseline~\cite{nguyen2025baselinesystemsevaluationmetrics}, the authors used this pre-trained M2D as the backbone feature extractor, along with a classification head that is trained on their own data from the DCASE challenge to perform audio tagging.

As for the TSE part, this baseline S5 system~\cite{nguyen2025baselinesystemsevaluationmetrics} uses a modified version of a previous TSE model~\cite{Kong2026UniversalSS}, which is a single-channel input and single-channel single-source output ResUNet model~\cite{Kong2021DecouplingMA}. This modified model accepts a multichannel audio mixture as input and predicts multichannel phase and magnitude masks, which are applied to the input. This mask prediction is conditioned using one-hot vectors obtained from the class labels generated by the tagger. These vectors are used to generate a parameter $\beta$, which is incorporated in the separator via a feature-wise linear modulation (FiLM)~\cite{Perez2017FiLMVR} of the network layers, resulting in a single-source multichannel output. Note that the original FiLM predicts two parameters $\gamma$ and $\beta$ based on the input label vecto~\cite{Perez2017FiLMVR}, but here only the additive parameter $\beta$ is used. To obtain the single-channel sources from this multichannel output, a 1 $\times$ 1 convolutional layer is applied. To perform separation, this model can then be applied multiple times during inference, each time using the class label corresponding to a given target source.

\subsection{Audio tagging models}
\label{sec:at models}
To design a lightweight S5 system, we explore two off-the-shelf audio tagging models. The first model is taken from the pretrained audio neural networks (PANNs) original paper~\cite{Kong2019PANNsLP}. This paper presents multiple model options for use as audio taggers, including convolutional neural networks (CNNs), convolutional residual networks (ResNets), MobileNets (MNs), and one-dimensional CNNs. These architectures were adapted by the authors to perform audio tagging on the AudioSet dataset. Among these available choices, we select the CNN6-based architecture due to its substantially lower parameter count compared to other architectures and its higher mean average precision (mAP) score on AudioSet~\cite{Kong2019PANNsLP}. The CNN6 backbone consists of stacked convolutional layers and a final fully-connected layer. For audio tagging, a linear classifier head is added, followed by a sigmoid activation function.

The second tagging model we use is a Dynamic MobileNet (DyMN)~\cite{Schmid2023DynamicCN}, which incorporates components such as dynamic convolutions~\cite{Chen2019DynamicCA} and coordinate attention~\cite{Hou2021CoordinateAF} into a global network design resembling the MN architecture~\cite{Howard2019SearchingFM}. The authors have proposed multiple versions of the DyMNs by varying the width multiplier $\alpha$ as mentioned in their paper. We select the smallest DyMN with $\alpha = 0.4$ for our experiments. The DyMNs have also been trained on AudioSet. We chose DyMN as a candidate for S5 audio tagger because it outperformed certain state-of-the-art (SOTA) audio tagging models in terms of mAP score, decreased complexity, and robustness exhibited towards varied downstream tasks~\cite{Schmid2023DynamicCN}.

\subsection{TSE models}
\label{sec:methodo-sep}

\begin{figure}[t]
    \centering
    \includegraphics[width=\columnwidth]{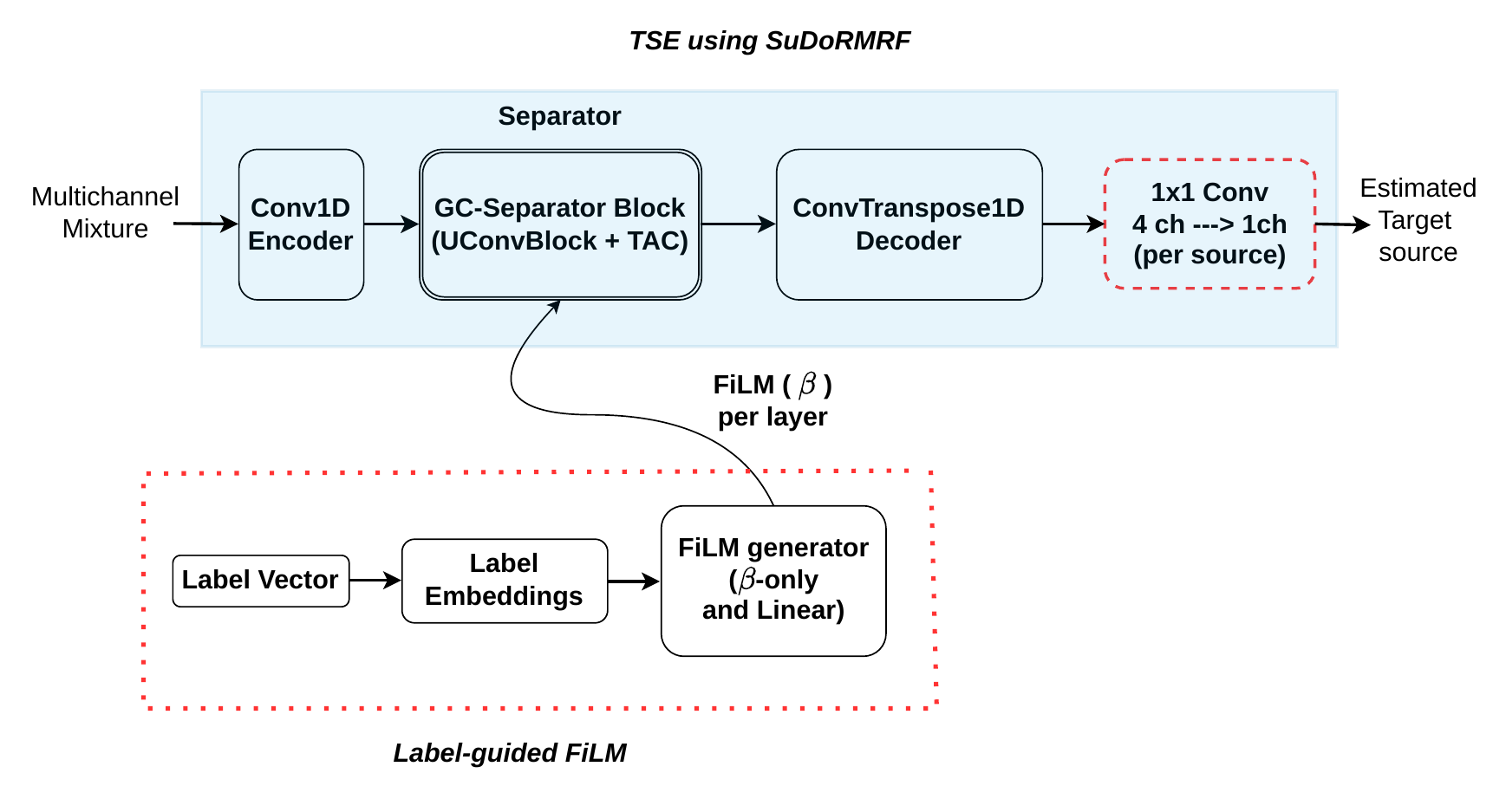}
    \vspace{-1.5em}
    \caption{Overview of the label-queried SuDoRMRF separator, with the modifications highlighted in red outlines.}
    \label{fig:separator}
    \vspace{-1em}
\end{figure}

As a lighter TSE model, we use a convolutional architecture~\cite{tzinis2022compute} that performs successive downsampling and resampling of multi-resolution features (SuDoRMRF). We chose this model due to its better performance against the SOTA models on various speech and non-speech source separation tasks. Specifically, out of several available versions of SuDoRMRF, we chose the group communications variant due to its high performance and extremely lightweight architecture~\cite{Luo2020UltraLightweightSS}. This model is an encoder-decoder style architecture that uses a transform-average-concatenate (TAC) module to communicate information between different groups, as displayed in Fig.~\ref{fig:separator}.
All these components contain different types of convolutional blocks. To increase or decrease model complexity, we can vary the number of U-ConvBlocks~\cite{tzinis2022compute}. SuDoRMRF's original implementation uses a single-channel audio mixture for separation and outputs single-channel source estimates.

To incorporate this TSE model within an S5 system, we modify it similarly to the way the ResUNet baseline was adapted (see Section~\ref{sec:methodo-baseline}), as illustrated in Fig.~\ref{fig:separator}. We adapt the first layer of the SuDoRMRF model so that it accepts a multichannel audio mixture as input. For the output, to obtain single-channel estimates from their multichannel counterparts, we apply a 1 $\times$ 1 convolutional layer as a projector, which reduces the number of channels from 4 to 1. Next, we incorporate the class label information of the target source via a FiLM module in the separator, following the baseline (\textit{cf}. Section~\ref{sec:methodo-baseline}). For each target source, the tagger outputs the class label as a one-hot vector. This parameter is created using a linear layer and is a vector whose dimensions depend on the number of input channels at the layer it is applied to. To reduce model complexity and computation, we only applied FiLM to the separator, and not to the encoder and decoder.

\section{Experimental Protocol}
\label{sec:exp}

\subsection{Dataset}
\label{sec:exp-dataset}

We use the development dataset designed for task 4 of the DCASE 2025 challenge, which contains sound events from 18 different classes~\cite{yasuda_2025_15117227}. To enable greater variability in the generated audio mixtures, we generate data on-the-fly for training and validation, similarly to the task's baseline system~\cite{nguyen2025baselinesystemsevaluationmetrics}.
 

We utilize the Spatial Scaper library~\cite{Romn2024SpatialSA} to generate 10-second-long audio mixtures at a sampling frequency of 32 kHz. Each generated mixture contains 1 to 3 target sound sources, with up to 3 overlapping sources. We set the default dataset length to $25,000$ samples per epoch during training, which corresponds to one fourth of the number of samples per epoch used in training the original baseline. This was done to reduce memory issues and make the training faster at each epoch. For validation after each training epoch, $544$ mixtures were generated. For the test set, we use the development dataset, which consists of 1500 mixtures.

\subsection{Training}

\subsubsection{Audio tagging models}

The lightweight audio tagging models mentioned in Section~\ref{sec:at models} were mostly trained on the AudioSet dataset with 527 classes, which is different from our dataset with 18 classes. Therefore, we instantiate a new head on top of the backbone extractor, and we then consider two training strategies. First, we train only the classifier head on our data, using the backbones from the original implementations. Using a pretrained backbone feature extractor allows for leveraging the knowledge from AudioSet while retraining the head accounts for the mismatch in the number of classes. Alternatively, we train the full model from scratch on our data.

As the loss function for training, we use the binary cross-entropy (BCE) loss. We train models with a batch size of 32 and for a maximum of 100 epochs. 
Since we do not use the multichannel information from the inputs to the S5 system for tagging models, we set the reference channel to 0, i.e., the omnidirectional channel. For both the CNN6 and the DyMN models, we use both training strategies as described above. Models for which we retrain only the head are denoted as CNN6-head and DyMN-head (for clarity, we use the same notation for the M2D baseline, therein denoted M2D-head). The models for which we retrain the full model from scratch are denoted as CNN6-full and DyMN-full.



\subsubsection{TSE models}

We train TSE models using oracle class labels for each source present in the mixture. We consider two versions of SuDoRMRF using either 16 or 4 U-ConvBlocks, and denoted as SuDoRMRF-16 and SuDoRMRF-4, respectively. We used the negative signal-to-distortion (SDR) as the loss function, and training is conducted for a maximum of 100 epochs with a batch size of 5.

\subsubsection{Baseline}

For a fair comparison, we retrain the baseline S5 system using our dataset described in Section~\ref{sec:exp-dataset}, but for reference we also report the original model (i.e., trained using a larger dataset) performance. These will be referred to as the \textit{retrained} and \textit{original} baselines, respectively, and the latter's sub-components (tagger and TSE) will be indicated with an asterisk (*).

\subsection{Evaluation metrics}

For evaluating the tagging and TSE models individually, we use variants of the class prediction accuracy and SDR that are specifically tailored for S5 evaluation~\cite{nguyen2025baselinesystemsevaluationmetrics}. For tagging we consider the \textit{mixture} accuracy, i.e., a prediction is counted as correct only if the labels for all the individual sources in the mixture are correct, as customary for S5 evaluation. For separation quality, we consider the class-aware SDR (CA-SDR), that was introduced by Nguyen et al. in~\cite{nguyen2025baselinesystemsevaluationmetrics}. This metric aligns the sources for SDR calculation in accordance with the class labels assigned to them by the tagger, i.e., it calculates the SDR only between true positive sources and assigns a zero SDR value to false positive and false negative ones. For our experiments, we use the official task metric for DCASE task 4 based on CA-SDR, i.e., CA-SDR improvement (CA-SDRi) in comparison to the mixture. A higher CA-SDRi value indicates better system performance. Finally, to evaluate the computational cost of each model, we report their number of parameters, in millions, and the number of multiply-accumulate operations (MACs), in billions.

\section{Results and Discussion}
\label{sec:r&d}

In this section, we present the experimental results. We first discuss the performance of alternative lightweight tagger and separator models individually, and then evaluate the overall S5 systems.

\subsection{Taggers performance}
\label{sec:results-tagger}

In Table~\ref{tab:tagger}, we present the performance of different audio taggers.

\begin{table}[t]
\centering
\caption{Performance of tagging models (mixture accuracy, in $\%$), and their sizes and MACs.}
\smallskip
\renewcommand{\arraystretch}{1.5}
\label{tab:tagger}

\begin{tabularx}{\linewidth}{
    l|
    >{\centering\arraybackslash}X|
    >{\centering\arraybackslash}X|
    >{\centering\arraybackslash}X}
    \hline
    \textbf{Tagger} 
    & \textbf{Mixture acc.} 
    & \textbf{Params. (M)} 
    & \textbf{MACs (B)} \\
    \hline
    M2D-head*   & 59.80 & 85.27 & 26.50 \\
    M2D-head    & 45.47 & 85.27 & 26.50 \\
    CNN6-head   & 19.73 &  4.58 &  6.20 \\
    CNN6-full   & 29.20 &  4.58 &  6.20 \\
    DyMN-head   & 37.13 &  0.80 &  0.03 \\
    DyMN-full   & 53.87 &  0.80 &  0.03 \\
    \hline
\end{tabularx}
\end{table}

We observe that the mixture accuracy drops by approximately 15\% between the original baseline tagger (M2D-head*) and the retrained baseline tagger (M2D-head). This is attributed to the fact that we used a smaller number of samples per epoch for retraining M2D, which is a large transformer-based model that is well-equipped to utilize larger amounts of data to its advantage. We also observe that the mixture accuracy drops for the lighter CNN6 model variants and DyMN-head when compared to the M2D-head tagger. This accuracy drop in general was to be expected, given the corresponding drop in the number of parameters. However, the DyMN-full model outperforms this model by almost 9\% accuracy and turns out to be the best performing tagging model. The CNN6-head and DyMN-head models have a lower accuracy when compared to their full counterparts. This points to the fact that the knowledge from the pretrained backbone is not being utilized properly, and there is a need to fine-tune the backbone on our dataset.

Overall, it appears worthwhile to use DyMN-full as a light tagger, as it exhibits the lowest parameter count and MACs, while maintaining an accuracy that is close to that of the M2D-head* baseline. In what follows, we focus on the DyMN tagger given its superior performance and lower computational cost compared to CNN6.

\subsection{Separators performance}
\label{sec:results-separator}

In Table~\ref{tab:separator}, we present a comparison between the baseline separators and their lightweight counterparts in terms of their performance and size. Here we report performance in oracle conditions, that is, using true source labels. We observe that the SuDoRMRF separator models outperform not only the retrained baseline but also the original one. Also when compared to the baselines, there is a drop in the number of parameters (almost 30 times) and MACs (about 3 times). Given such results, SuDoRMRF appears as a viable alternative as a separator to create a lightweight S5 system. In what follows, we focus on the SuDoRMRF-16 variant, given its slighlty superior performance compared to SuDoRMRF-4.

\begin{table}[t]
\centering
\caption{Performance of TSE models using oracle labels (CA-SDRi in dB), and their sizes and MACs.}
\smallskip
\label{tab:separator}
\renewcommand{\arraystretch}{1.5}
\begin{tabularx}{\linewidth}{
    l|
    >{\centering\arraybackslash}X|
    >{\centering\arraybackslash}X|
    >{\centering\arraybackslash}X}
    \hline
    \textbf{Separator} 
    & \textbf{CA-SDRi} 
    & \textbf{Params. (M)} 
    & \textbf{MACs (B)} \\
    \hline
    ResUNet*        & 12.93  & 29.87 & 809.39 \\
    ResUNet         & 10.74 & 29.87 & 809.39 \\
    SuDoRMRF-16  &  13.61  &  1.91 & 329.87 \\
    SuDoRMRF-4   &  13.20 &  1.34 & 256.63 \\
    \hline
\end{tabularx}
\end{table}

\begin{table}[t]
\centering
\caption{Performance of S5 systems. Both the original and retrained baseline systems are reported in the first two rows for reference.}
\label{tab:S5_system}
\smallskip
\renewcommand{\arraystretch}{1.5}

\begin{tabularx}{\linewidth}{
    l|
    >{\centering\arraybackslash}X|
    >{\centering\arraybackslash}X|
    >{\centering\arraybackslash}X}
    \hline
    \textbf{S5 system} 
    & \textbf{CA-SDRi} 
    & \textbf{Params. (M)} 
    & \textbf{MACs (B)} \\
    \hline

    M2D-head* + ResUNet*    & 11.03 & 115.38 & 835.89 \\
    M2D-head + ResUNet      &  7.89 & 115.38 & 835.89 \\
    \hline
   M2D-head* + SuDoRMRF-16 & $11.01$ & 87.43 & 356.37 \\
   DyMN-full + ResUNet* & 10.36 & 31.57 & 809.42 \\ 
    DyMN-full + SuDoRMRF-16 & 10.73 &   3.62 & 329.90 \\
    \hline
   CNN6-head + SuDoRMRF-16 & $5.00$ & 6.50 & 336.10\\
    \hline
\end{tabularx}
\end{table}

\subsection{S5 systems performance}

Table~\ref{tab:S5_system} presents the performance of various S5 systems. First, we observe that retraining the baseline with our smaller-size dataset yields a significant performance drop, which is in line with our conclusions from the previous sections. Interestingly, we observe that combining the large M2D-head* with SudoRMRF-16 maintains performance compared to the original ResUNet* baseline separator, which confirms the potential of this lightweight separator. Conversely, replacing the baseline tagger with our lightweight DyMN-full while keeping the large ResUNet* separator,results in a CA-SDRi drop of approx 0.7 dB: even though this performance drop is more important, it comes with a more substantial reduction in terms of model complexity. Lastly, our proposed lightweight alternative made up of DyMN-full and SudoRMRF-16 not only surpasses the retrained baseline but also performs on par with the original one. Indeed, this system reduces the number of parameters by approximately $97\%$ and the number of MACs by 2.5 times, while experiencing moderate decreases in mixture accuracy and CA-SDRi of about~$6\%$ and~$0.45$ dB, respectively.

\begin{figure}[t]
    \centering
    \includegraphics[width=\columnwidth]{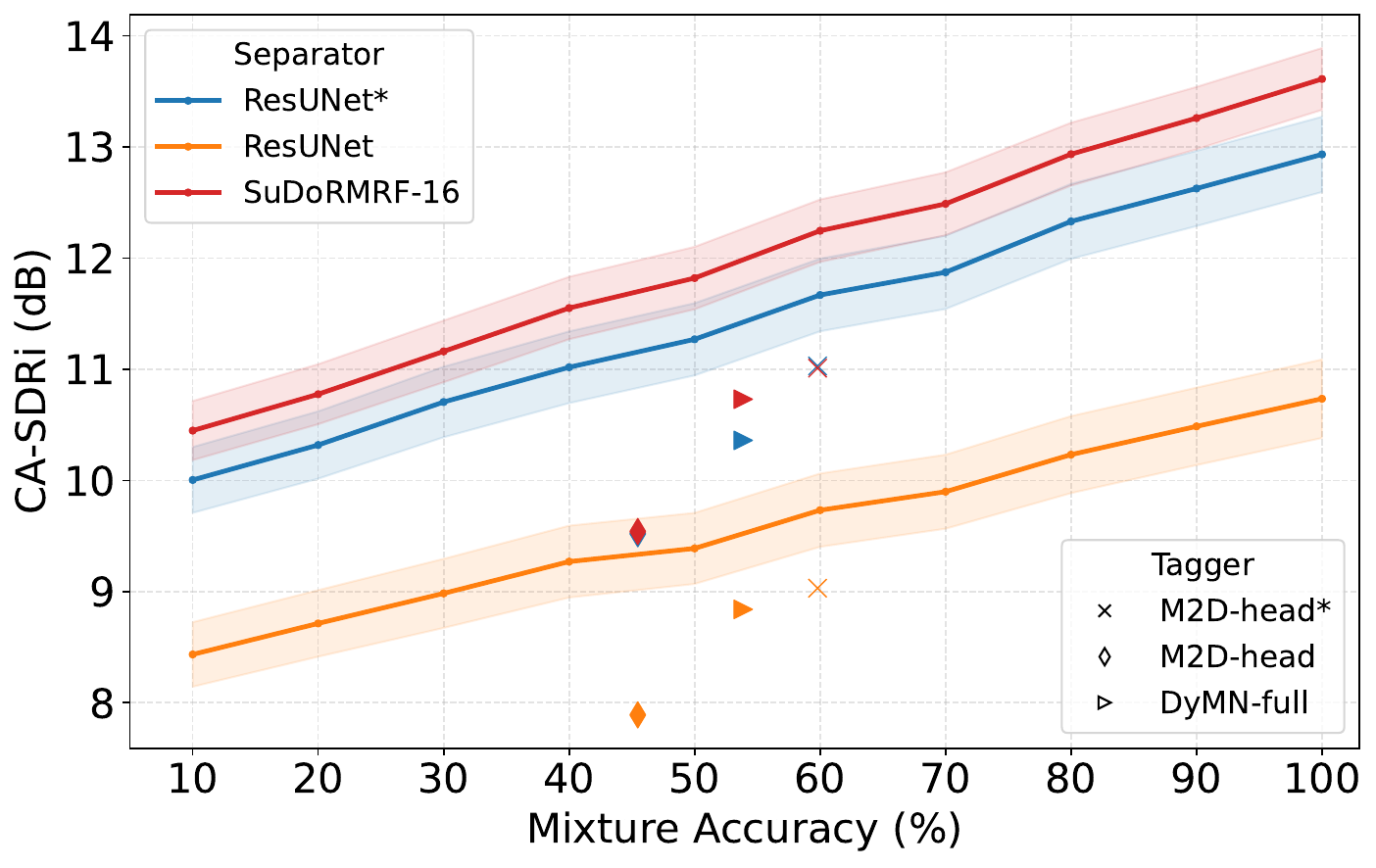}
    \vspace{-1.5em}
    \caption{Comparing CA-SDRi for different separators as a function of tagger accuracy.}
    \label{fig:separation_vs_accuracy}
\end{figure}

In the last row of Table~\ref{tab:S5_system}, we report the performance of the poorly-performing CNN6 tagger when used in conjunction with SudoRMRF-16. We observe a significant 5 dB CA-SDRi drop when compared to using DyMN, which outlines the importance of a properly performing tagger for general S5 performance. To further investigate this compound impact of the tagger and separator, we conduct an additional experiment where we artificially introduce errors from oracle labels, in order to track the impact of mixture accuracy onto CA-SDRi. Results are displayed in Figure~\ref{fig:separation_vs_accuracy}. We notice that each S5 system's performance increases almost linearly with the increase in the tagger accuracy. Besides, we see that the SuDoRMRF model outperforms the baselines at all accuracy levels. We also plot the performance of individual S5 systems with different taggers. In most cases, labeling errors introduced artificially yields to the same trend as using real taggers (i.e., SuDoRMRF-16 outperforms ResUNet variants). One exception is the M2D-head* tagger, for which similar results are obtained using SuDoRMRF-16 and ResUNet. 

A noticeable observation is that for a given separator, the CA-SDRi is higher under controlled labeling errors than when using a real tagger. One explanation is that we introduce these errors only in the form of label \textit{substitutions}, whereas an actual tagger can also make label \textit{deletion} or \textit{swapping} - although the latter has no impact on the CA-SDRi, as investigated in a prior study~\cite{mishra2026metricanalysisspatialsemantic}. Thus, there is more semantic content available for the separator to perform TSE when artificially introducing labeling errors. Besides, here we only introduced one substitution per mixture, while a real tagger might introduce more, which would result in a lower CA-SDRi while the mixture accuracy would be the same. Studying the behavior of the separators in such a fine-grain setup where we control various possible sources and number of labeling errors is outside the scope of this study, albeit it opens interesting research directions for future work.

\section{Conclusion}
\label{sec:conc}

In this paper, we compared multiple S5 systems created by using lighter audio tagging and separation models. We modified the SuDoRMRF separation model to perform TSE in accordance with the S5 baseline. We found that the DyMN and SuDoRMRF models serve as great alternatives to the heavier M2D-head and ResUNet baseline models. We highlight the fact that for these S5 systems, the number of parameters and MACs of the tagger can be greatly reduced, approximately 105 times, without adversely affecting the accuracy. For the separator, the reduction in the number of parameters can only be as large as 15 times, but the number of MACs is not reduced by a large multiplier. These reductions in the number of parameters and MACs of the proposed S5 systems still fall short for certain edge applications~\cite{Fedorov2020TinyLSTMsEN, benslimane2026rttangorealtimedistributedbinaural}. Therefore, in future work, we will research various other techniques, such as knowledge distillation, quantization, or pruning, to further reduce these systems' size.


\section{Acknowledgments}
This research was carried out with the support of the French National Research Agency as part of the CONFLUENCE project number \text{ANR-23-EDIA-0003}. Experiments presented in this paper were carried out using the \href{https://www.grid5000.fr}{Grid'5000} testbed, supported by a scientific interest group hosted by Inria and including CNRS, RENATER, and several universities, as well as other organizations.

\bibliographystyle{IEEEtran}

\bibliography{strings,refs}

\end{document}